# The Haber Process Made Efficient by Hydroxylated Graphene

Vitaly V. Chaban[1,2] and Oleg V. Prezhdo[2]

[1] Instituto de Ciência e Tecnologia, Universidade Federal de São Paulo, 12231-280, São José dos Campos, SP, Brazil

[2] Department of Chemistry, University of Southern California, Los Angeles, CA 90089, USA

**Abstract**. The Haber-Bosch process is the main industrial method for producing ammonia from diatomic nitrogen and hydrogen. Very demanding energetically, it uses an iron catalyst, and requires high temperature and pressure. Any improvement of the Haber process will have an extreme scientific and economic impact. We report a significant increase of ammonia production using hydroxylated graphene. Exploiting the polarity difference between $N_2/H_2$ and $NH_3$, as well as the universal proton acceptor behavior of $NH_3$, we demonstrate a strong shift of the equilibrium of the Haber-Bosch process towards ammonia. Hydroxylated graphene provides the polar environment favoring the forward reaction, and remain stable under the investigated thermodynamic conditions. Ca. 50 kJ mol$^{-1}$ enthalpy gain and ca. 60-70 kJ mol$^{-1}$ free energy gain are achieved at 298-1300 K and 1-1000 bar, strongly shifting the reaction equilibrium towards the product. A clear microscopic interpretation of the observed phenomenon is given using electronic structure calculations and real-time reactive simulations. The demonstrated principle can be applied with many polar groups functionalizing a substrate with a high surface area, provided that the system is chemically inert to $H_2$, $N_2$ and $NH_3$. The modified Haber-Bosch process is of significant importance to the chemical industry, since it provides a substantial increase of the reaction yield while decreasing the temperature and pressure, thereby, reducing the cost.

# 1. Introduction

Binding atmospheric nitrogen is in the core of chemical technology and many life processes. Nitrogen is abundant in the air. Certain unique bacteria use the nitrogenase enzyme to bind and consume nitrogen in small amounts. An artificial fixation of nitrogen still remains a great challenge to the civilization. Arguably, the development of the Haber-Bosch process a century ago contributed more to the human well-being in the past century than the inventions of the aeroplane, television, nuclear energy, space shuttles, and computers. The Haber-Bosch process served as a detonator of the human population explosion, enabling more than a four-fold population growth within the last few generations, since the production of ammonia is directly related to the production of food.[1]

The Haber-Bosch process consumes high amounts of energy, and therefore, it is economically demanding. Depending on a particular technology, the synthesis takes place at 700-900 K and 200-400 bar. Maintaining both high temperature and high pressure for the ammonia synthesis worldwide consumes one-two percent of all energy generated by humans.[2] A fundamental scientific finding or an engineering development that can either improve the reaction yield and the catalyst performance, or adjust the synthesis conditions, will have great global significance.

Despite the extreme simplicity of the chemical reaction, $N_2 + 3H_2 \leftrightarrow 2NH_3$, the process appears highly resistant to technological optimization. Based on the purely thermodynamic consideration, ammonia should be spontaneously synthesized at room conditions. In reality, the rate of this reaction is so low that it can be scarcely observed. The primary reason is that the ternary nitrogen-nitrogen bond, 942 kJ mol$^{-1}$, must be broken prior to binding three hydrogen atoms. This slowest step determines the overall reaction rate. Metal catalysts are employed to create transition complexes with molecular nitrogen fostering its dissociation.[3-6]

The pursuit continues. Many catalysts for the ammonia synthesis have been introduced since the Haber-Bosch process has been implemented on an industrial scale. Iron, cobalt, ruthenium, molybdenum, cerium, and carbon containing compounds with admixtures of potassium, calcium, aluminum, and silicon have been proposed at different times.[7] Many of these catalysts themselves require high temperatures, ca. 700 K, to act efficiently in relation to $N_2$. Dinitrogen is expected to undergo dissociative adsorption on the catalyst surface to provide N*. Afterwards, each N* reacts with 1.5×$H_2$ to produce $NH_3$. Ultimately, $NH_3$ leaves the catalyst surface under the pressure of other nitrogen molecules. The tools employed to cleave the nitrogen-nitrogen triple bond can be divided into three groups: (1) reductive catalysts, which provide electrons to $N_2$ (this type of activation occurs in the Haber-Bosch process); (2) oxidative catalysts, which take electrons away from $N_2$ (no such catalysts were developed thus far, since the first ionization energy of $N_2$ is high; (3) extreme conditions, such as high voltage, plasma treatment, and photo-induced electron excitation: $N_2 \rightarrow 2N^*$, $N_2 - e^- \rightarrow 2N^+$. The last method consumes excessive amounts of energy, preventing its cost-efficient industrial implementation. Several novel catalysts allow one to adjust somewhat the reaction temperature and pressure, e.g. Ru/C catalysts;[8,9] $Ru/C_{12}A_7$:$e^-$ with the highly electron donating $12CaO \cdot 7Al_2O_3$ support;[10] metal complexes inducing hydrogenation at ambient conditions;[11] the proton-conducting cell reactor,[12] photo-catalysts,[13] etc.

While the main research efforts are concentrated on the development of catalysts able to increase the rate[14,15] of $N_2$ cleavage, another fundamental option to adjust ammonia yield has been largely omitted. The $N_2 + 3H_2 \leftrightarrow 2NH_3$ process constitutes a relatively non-abundant type of chemical reaction, in which two non-polar particles form one strongly polar particle. Indeed, $N_2$ and $H_2$ have no dipole moments, while the dipole moment of $NH_3$ is 1.4 D. This is much larger than the dipole moment of phosphine, 0.58 D, the phosphorus analog of ammonia. Furthermore, ammonia is a universal proton acceptor. This property is rarely utilized during its numerous reactions with organic compounds. If the newly formed $NH_3$ molecules are provided an opportunity to interact with a proton-donating group, the reaction thermodynamics can change significantly, strongly affecting the equilibrium constant and enabling higher $NH_3$ yields.

We use ab initio quantum chemistry[16] and reactive chemical dynamics[17] simulations to demonstrate up to a 9% increase of the ammonia yield in the Haber-Bosch process using hydroxylated graphene quantum dots (GQDs, Figure 1). Hydroxylated graphene provides the polar environment shifting the equilibrium towards the product, while at the same time remaining intact under the investigated thermodynamic conditions, including 298-1300 K and 1-1000 bar. The closely related alcohols, $C_xH_yOH$, do not react with $NH_3$ (K<<1), since the $NH_3$ leaving group, which is the conjugate base of $NH_4^+$, is much stronger than the $OH^-$ leaving group, which is the conjugate base of $H_2O$. Despite a strong interaction, the aminated GQD is not formed, leaving $NH_3$ intact, as necessary. Nonpolar dinitrogen and dihydrogen remain indifferent to the hydroxylated GQD, and the thermodynamic equilibrium is shifted from the reactants to the product. 50 kJ mol$^{-1}$ enthalpy and 60-70 kJ mol$^{-1}$ free energy gains are achieved.

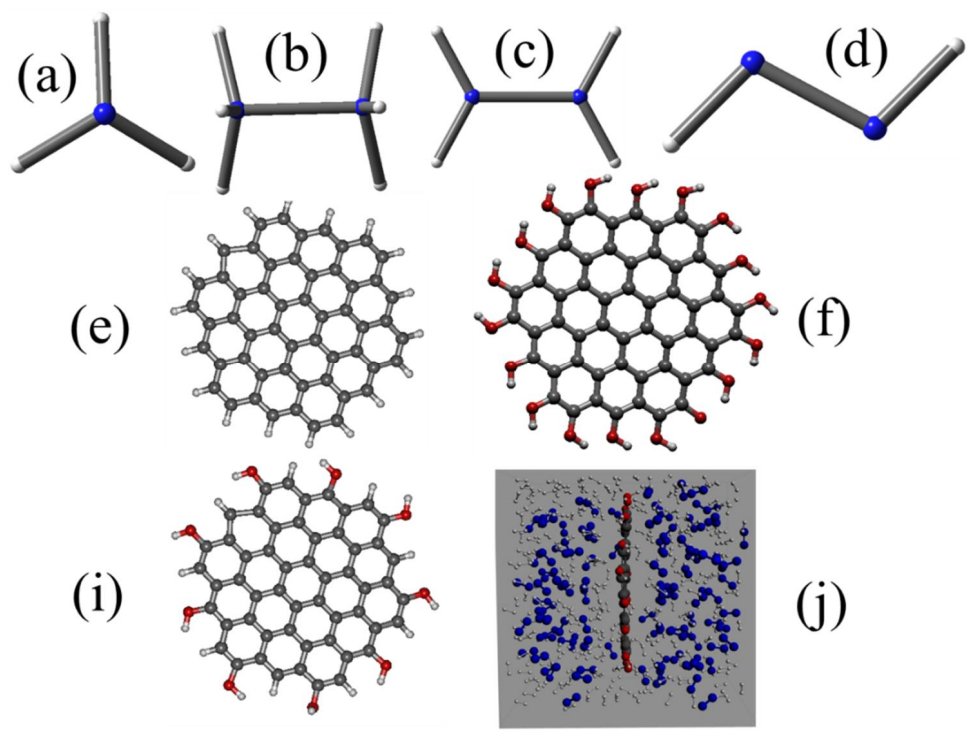

**Figure 1.** Molecules participating in the synthesis of ammonia: (a) ammonia $NH_3$; (b) diammonia $N_2H_6$; (c) hydrazine $N_2H_4$; (d) diazene $N_2H_2$; (e) GQD; (f) hydroxylated GQD-I; (i) hydroxylated GQD-II; (j) elementary cell for ammonia synthesis.

## 2. Theory

The enhancement of the yield of the ammonia synthesis via the Haber process has been demonstrated by two independent techniques. The thermodynamics of the process have been investigated by the highly accurate ab initio Gaussian-4 (G4) theory. The kinetic details of the chemical reaction have been investigated further using reactive molecular dynamics (RMD).

**2.1 Thermodynamics**

The recent G4 theory offers an effective chemical accuracy of the thermodynamic quantities in the gas phase, such as enthalpy, entropy, Gibbs free energy, heat of formation, heat capacity, etc. G4 is based on a series of energy evaluations, as well as a thorough geometry optimization at the beginning.[18] The G4MP2 method is a variation of G4, wherein MP3 and MP4 levels of theory are avoided to decrease the computational cost, in particular the computer memory requirement. G4MP2 still includes an effect of electron correlation,[18] since electron correlation is important for deriving accurate molecular partition functions. The average absolute deviation from the experimental data, evaluated over a large test set (454 energies) equals to 4.3 kJ mol$^{-1}$. This accuracy is considered good for chemical applications, exceeding that of the earlier methods. The G4MP2 method is still expensive and can hardly be applied to large (and periodic) systems, such as nanotubes, graphene, quantum dots and biomolecules. The implementations of G4 and G4MP2 are available in the GAMESS quantum chemistry suite.[19]

**2.2 Reactive Molecular Dynamics**

The reported simulations of the ammonia synthesis reaction from molecular nitrogen and molecular hydrogen were performed by the RMD method. RMD employs quantum chemistry (QC) based reactive force fields (ReaxFF).[17,20] The methodology was applied previously with considerable success to address complicated problems in organic chemistry and catalysis.

ReaxFF provides a realistic description of reactive potential surfaces for many-component systems. It allows one to reproduce bond formation and breaking without optimizing the corresponding wave functions. Simulations including a few thousands of interaction centers are within reach of RMD. The instantaneous electrostatic point-charge residing on each atom is evaluated from the instantaneous electrostatic field. Interaction between two charged particles is represented as a shielded Coulomb potential. The instantaneous valence force and interaction energy are determined through the instantaneous bond order. All the enumerated interaction energy functions were parametrized with respect to the accurate QC energy scans involving all applicable types of atom hybridization and all types of expected bond-breaking and bond-formation processes.[17,20] The well-known pairwise van der Waals energy component describes short-range electron-electron repulsion and dispersion attraction ($E \sim -r^{-6}$). One of its roles is to preserve atom radii during a reaction. The required components of the model reactive Hamiltonian were derived from multiple density functional theory calculations. Further details on the development and implementation of reactive simulations are available elsewhere.[17,20] The reported simulations of the ammonia synthesis were carried out using the implementation of the method in ADF2013 (scm.com).

## 3. Results

### 3.1 Thermodynamics

Figure 2 provides a thermodynamic analysis of the ammonia synthesis reaction with and without the hydroxylated GQD for a wide range of thermodynamic conditions. To allow application of an accurate ab initio method to obtain molecular partition functions, we used 1,3,5-trihydroxybenzene as a simplified model of the hydroxylated GQD. This simplification is reasonable, since the coordination of $NH_3$ occurs at the hydrogen atoms of the OH groups exhibiting the proton acceptor behavior, whereas the nonpolar GQD core is inert. A few geometries of different size GQDs were optimized to derive this conclusion. The established G4MP2 method was used to obtain the thermodynamic quantities by a sequence of the predefined procedures.[18,21] The equilibrium N(NH$_3$)-H(OH) distance amounts to 1.87 Å, which significantly exceeds the 1.02 Å N-H bond length in NH$_3$. The distance of 1.87 Å qualifies for a hydrogen bond. The angles involving the hydrogen bond, H-N-H, are 105, 115, 116°, while the H-N-H angles in the coordinated NH$_3$ are 107 Å each. The hydrogen bond brings a significant energy gain favoring the forward reaction, ca. 50 kJ mol$^{-1}$ of enthalpy and ca. 60-70 kJ mol$^{-1}$ of free energy.

Generation of NH$_3$ from H$_2$ and N$_2$ is exothermic. According to the Châtelier's principle, temperature increase is unfavorable, since it shifts the equilibrium towards the reactants. At the same time, the reaction is too slow at room temperature. Formation of NH$_3$ decreases the number of gas molecules in the reaction mixture, and hence pressure, by a factor of two. Thus, an increased pressure is favorable for the reaction product. These considerations are confirmed by computational thermodynamics (Figure 2), both with and without the hydroxylated GQD.

Low temperature and high pressure constitute the ideal thermodynamic conditions for the NH$_3$ synthesis. Note that generation of high pressure is expensive in many aspects, especially on the industrial scale. Pipes and reaction vessels have to be strengthened. High quality valves are needed. Running pumps and compressors takes considerable amount of energy. Working at 200 bar requires extra safety precautions. A balance between reaction yield and operation expenses has to be found for the synthesis to be cost efficient. The industry considers a single pass to be efficient if it yields ca. 15% of the product. Multiple cycles are conducted to achieve an acceptable nitrogen conversion degree, above 90%. If the equilibrium in every cycle is shifted to the products, as suggested by the thermodynamics analysis of Figures 2 and 3, the resulting performance gain will be very significant.

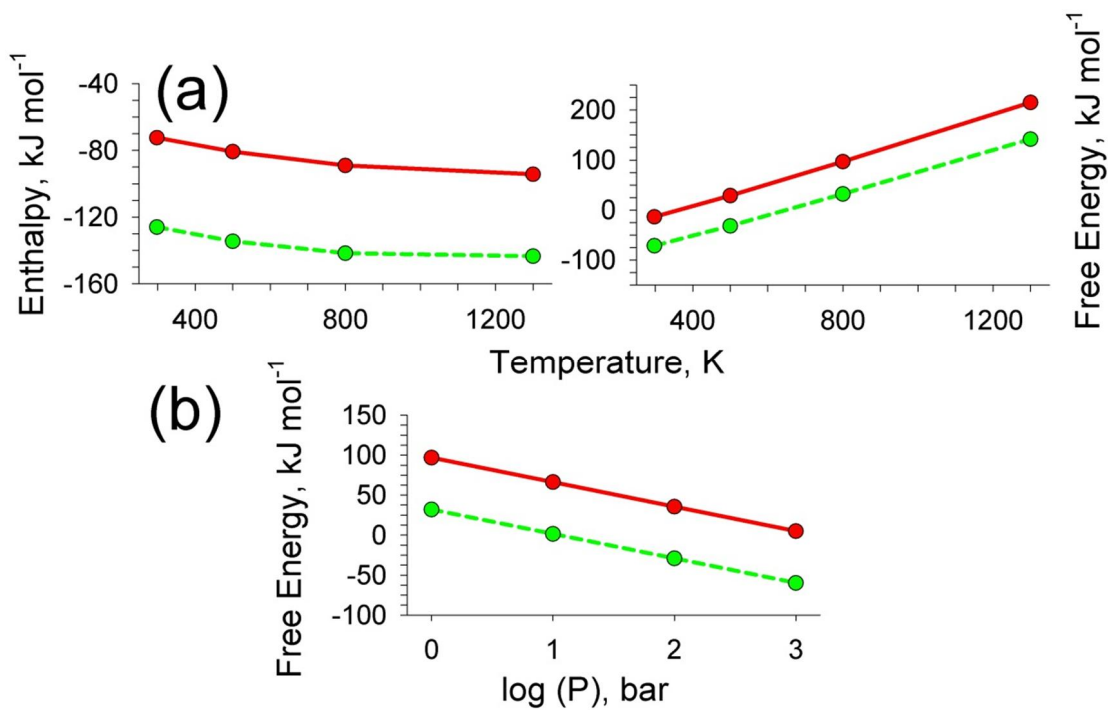

**Figure 2.** Thermodynamic potentials of the ammonia synthesis reaction vs. temperature and pressure without (red solid line) and with (green dashed line) the hydroxylated GQD. Figure 3 investigates the entropic contribution. (a) Temperature was changed 300 to 1300 K at 1 bar; (b) Pressure was changed 1 to 1 000 bar at 800 K.

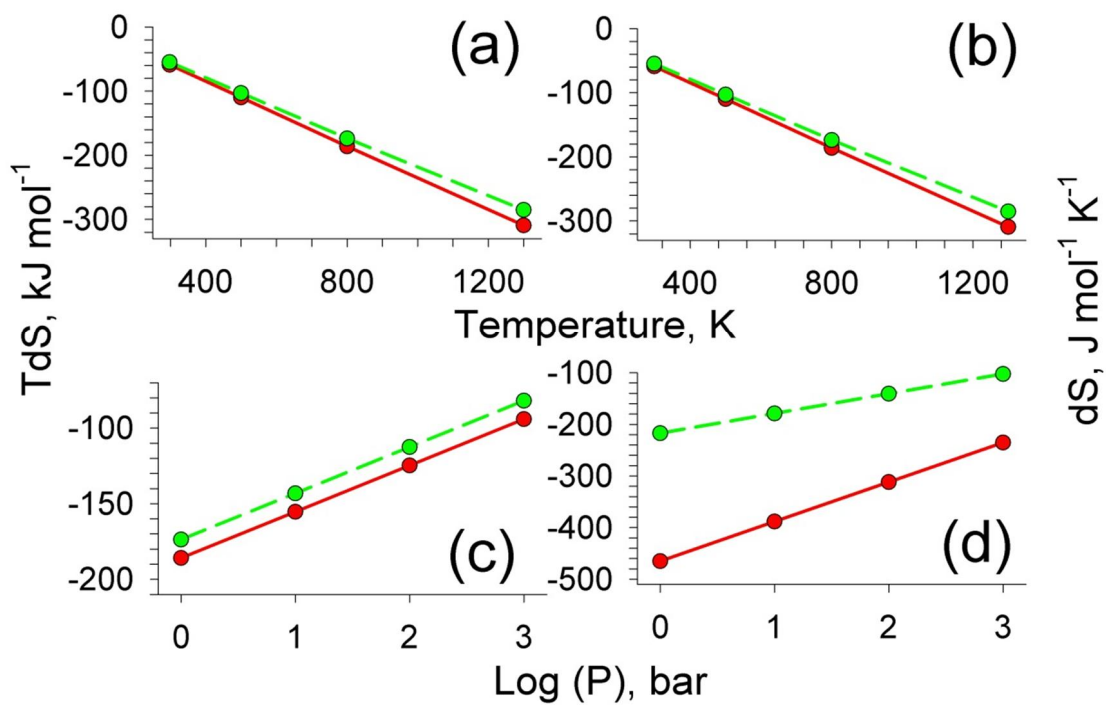

**Figure 3.** Entropic factor and entropy as functions of temperature and pressure for the ammonia synthesis from the elements: in the bulk mixture (red solid line); with the hydroxylated GQD (green dashed line). The reaction is strongly entropically forbidden, especially at high temperatures.

**3.2 Reactive Molecular Dynamics**

To support our thermodynamic predictions with realistic, many-particle systems, we carried out RMD simulations using the force field model provided by van Duin and coworkers.[17,20] The parametrization of the method is based on reproducing a potential energy surface relevant to the chemical reactions in question using multiple density functional theory calculations. Each simulated system consisted of ca. 1000 atoms, including 100 dinitrogen and 300 dihydrogen molecules (Figure 1j). The $N_2 + 3H_2 \leftrightarrow 2NH_3$ reaction is very slow, and establishing thermodynamic equilibrium takes macroscopic times. Therefore, the RMD simulations were conducted at 1500 K with the reactant mixture density of 710 kg m$^{-3}$. If needed, the results can be extrapolated to the conditions of interest. The equations-of-motion were propagated with a 0.0001 ps time-step, and each system was simulated during 100 ps, achieving a sampling of 1 000 000 molecular configurations.

Nitrogen molecules decompose quickly at the elevated temperature (Figure 4). After 60 ps of the RMD simulation, their content is below 10 mol%, while over 20 mol% of hydrogen molecules are still available. Significant amounts of diazene and hydrazine are generated. The $N_2H_2$ content reaches maximum at around 20 ps and then slowly decreases. In turn, the content of $N_2H_4$ vs. time exhibits no maximum. In most cases, the difference between the systems with and without the hydroxylated GQD can be seen. Note that the reaction is simulated without the metallic catalyst, for simplicity, since the catalyst does not impact the reaction equilibrium.

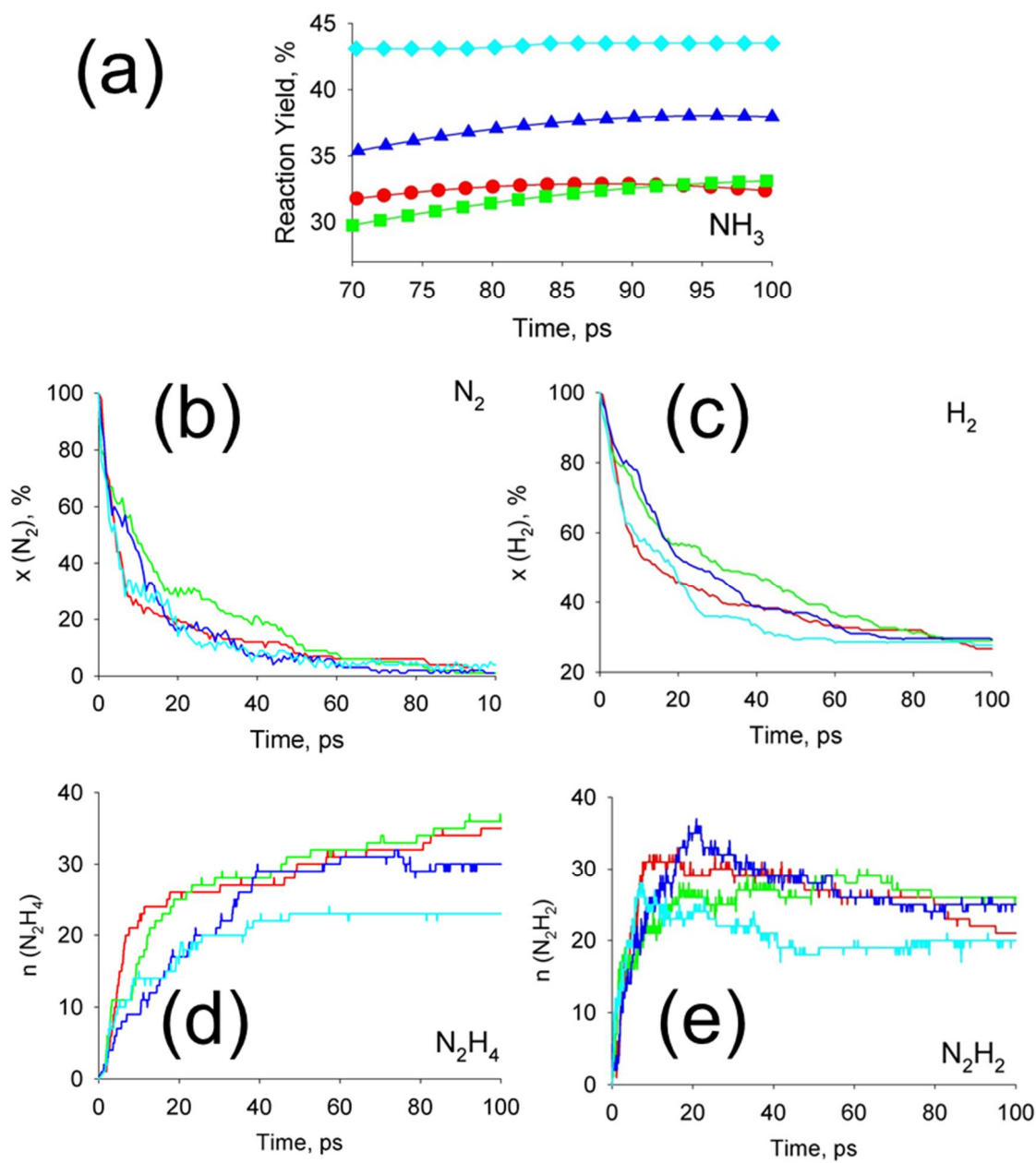

**Figure 4.** Generation of products and decay of reactants in the course of the ammonia synthesis reaction: (a) $NH_3$; (b) $N_2$; (c) $H_2$; (d) hydrazine $N_2H_4$; (e) diazene $N_2H_2$. The bulk mixture of the reactants without GQDs is red; with pristine GQD (Figure 1e) is green; with hydroxylated GQD (Figure 1f) is dark blue; with two hydroxylated GQDs is blue.

Real-time visual examination and statistical analysis, such as that presented in Figure 4, allows us to assign the following mechanism to the simulated reaction. Dihydrogen attaches to dinitrogen forming various chains of low stability, such as …-H-H-N=N-H-H-… . The triple bond in $N_2$ is hereby substituted by the double bond. These chains quickly break down generating $N_2H_2$. According to Figure 4, $N_2H_2$ molecules are formed already during the initial simulation steps. An additional $H_2$ molecule binds to $N_2H_2$ shortly after, to generate $N_2H_4$, which finally consumes a third $H_2$ molecule and transforms into $N_2H_6$. Diammonia $N_2H_6$ is unstable and can be observed only for a few femtoseconds before it cleaves into two $NH_3$ molecules.

The presence of the hydroxylated GQD is not favorable for the last $N_2H_4+H_2 \rightarrow N_2H_6 \rightarrow 2NH_3$ transformation, since $N_2H_4$ is more polar, $\mu=1.85$ D, than $NH_3$, and it also interacts with the OH groups. However, this reaction is strongly favorable energetically, $\Delta G=-291$ kJ mol$^{-1}$, at ambient conditions. The overall $NH_3$ yield is dependent on the presence of the hydroxylated GQD (Figure 4a). It increases from 33 to 38% when a single hydroxylated GQD (Figure 1f,i) is provided and to 44% when two hydroxylated GQDs are provided. To isolate the impact of the GQD itself, we simulated a pristine GQD (Figure 1e) and observed the same yield, 33%, as in the case of the bulk mixture. Thus, the reaction equilibrium gets shifted due to the hydroxyl groups, while the GQD acts exclusively as a matrix/support. We also compared GQDs with higher (Figure 1f) and lower (Figure 1e) content of hydroxyl groups. It was observed that both GQDs provide comparable $NH_3$ yields, ca. 38%. We found that a small distance separating the hydroxyl groups, such as in Figure 1f, favors intra-molecular hydrogen bonding, being therefore inefficient. This conclusion is also confirmed by the RMD simulation of the two hydroxylated GQDs. When the distance between the hydroxylated GQDs was fixed to 10 Å, the yield was 44%. In turn, when they were allowed to approach one another, the yield dropped down to 35%, which is only modestly higher than that in the bulk mixture.

The theoretical threshold of the reaction yield can be calculated from the thermochemical results via the following formula, $K=\exp(-\Delta G/RT)$, where K is the equilibrium reaction constant, R is the gas constant; T is temperature; and $\Delta G$ is the free energy change. Compare, $\Delta G=+35.7$ kJ mol$^{-1}$ at 800 K and 100 bar (similar to the technological conditions) without the hydroxylated GQD, and $\Delta G=-29.1$ kJ mol$^{-1}$ with the hydroxylated GQD at the same temperature and pressure. Consequently, the corresponding equilibrium constants differ by more than a factor of 10 000.

The proposed method allows one to adjust reaction conditions appropriately. It does not substitute or compete with the metallic catalysts to cleave $N_2$. The metallic catalysts do not contain polarized hydrogen atoms, which would interact strongly with $NH_3$. Additionally, the topology of the metal catalysts is not optimal for hosting multiple hydroxyl groups. Graphene or GQDs can be decorated with multiple hydroxyl groups, which would be evenly distributed not only at the edge positions, but also throughout the surface. This material is easily fabricated from graphite oxide. The latter is obtained by acting with potassium permanganate on graphite, which subsequently exfoliates. It is a standard method to oxidize organic compounds. The hydroxylated graphene can be arranged in layers to maximize its working surface area. It is particularly beneficial in practice that the hydroxylated graphene does not chemically modify ammonia.

## 4. Conclusions

To recapitulate, we introduced a new method for controlling ammonia synthesis by exploiting differences in the polarities of the reactants and the product, and the strong proton acceptor ability of $NH_3$. We reported a significant performance increase, up to 9% in ammonia yield, at 1500 K and the density of 710 kg m$^{-3}$. The differences between the original Haber-Bosch process and the improved process generally depend on external conditions. The proposed method was justified by two independent theoretical approaches, providing strong support of the suggested technique. The advance allows one to decrease the number of passes/stages to obtain a desired degree of nitrogen conversion. The method also allows one to decrease the target temperature and pressure. It is particularly interesting to use the proposed method for shifting the thermodynamic equilibrium in conjunction with the novel catalysts,[10] which accelerate the kinetics. The topology of the functionalized GQD and graphene deserve further engineering in order to increase the product yield. The demonstrated principle is general and can be applied with many polar groups functionalizing a substrate with a high surface area, as long as the groups and the substrate are not chemically reactive with $H_2$, $N_2$ and $NH_3$. The achieved advance is expected to have great economic and societal impacts.